\documentclass[journal=NanoLetters,manuscript=article]{achemso}
\setkeys{acs}{keywords = true}

\usepackage[version=3]{mhchem} 
\usepackage{graphicx}
\usepackage{dcolumn}
\usepackage{multirow}
\usepackage{subfigure}
\usepackage{epsfig}
\usepackage{xcolor}
\usepackage{bm}
\usepackage{color}
\usepackage{amsmath}
\usepackage{amsfonts}
\usepackage{amssymb}
\usepackage{booktabs}
\usepackage{listings}
\usepackage{lmodern}  
\usepackage{microtype}
\usepackage{natmove}
\usepackage{natbib}

\author{Wei Xun}
\affiliation{State Key Laboratory for Mechanical Behavior of Materials, Center for Spintronics and Quantum System, School of Materials Science and Engineering, Xi'an Jiaotong University, Xi'an, Shaanxi, 710049, China}
\alsoaffiliation{Faculty of Electronic Information Engineering, Huaiyin Institute of Technology, Huaian 223003, China}
\author{Chao Wu}
\affiliation{State Key Laboratory for Mechanical Behavior of Materials, Center for Spintronics and Quantum System, School of Materials Science and Engineering, Xi'an Jiaotong University, Xi'an, Shaanxi, 710049, China}
\author{Hanbo Sun}
\affiliation{State Key Laboratory for Mechanical Behavior of Materials, Center for Spintronics and Quantum System, School of Materials Science and Engineering, Xi'an Jiaotong University, Xi'an, Shaanxi, 710049, China}
\author{Weixi Zhang}
\affiliation{Department of Physics and Electronic Engineering, Tongren University, Tongren 554300, China}
\author{Yin-Zhong Wu}
\affiliation{School of Physical Science and Technology, Suzhou University of Science and Technology, Suzhou 215009, China}
\author{Ping Li}
\email{pli@xjtu.edu.cn}
\affiliation{State Key Laboratory for Mechanical Behavior of Materials, Center for Spintronics and Quantum System, School of Materials Science and Engineering, Xi'an Jiaotong University, Xi'an, Shaanxi, 710049, China}
\alsoaffiliation{State Key Laboratory for Surface Physics and Department of Physics, Fudan University, Shanghai, 200433, China}

\title[An \textsf{achemso} demo]
{Coexisting Magnetism, Ferroelectric, and Ferrovalley Multiferroic in Stacking-Dependent Two-Dimensional Materials}


\begin{document}

\begin{abstract}
The two-dimensional (2D) multiferroic materials have widespread of application prospects in facilitating the integration and miniaturization of nanodevices. However, it is rarely coupling between the magnetic, ferroelectric, and ferrovalley in one 2D material. Here, we propose a mechanism for manipulating magnetism, ferroelectric, and valley polarization by interlayer sliding in 2D bilayer material. Monolayer GdI$_2$ exhibits a ferromagnetic semiconductor with the valley polarization up to 155.5 meV. More interestingly, the magnetism and valley polarization of bilayer GdI$_2$ can be strongly coupled by sliding ferroelectricity, appearing these tunable and reversible. In addition, we uncover the microscopic mechanism of magnetic phase transition by spin Hamiltonian and electron hopping between layers. Our findings offer a new direction for investigating 2D multiferroic in the implication for next-generation electronic, valleytronic, and spintronic devices.
\end{abstract}

Two-dimensional (2D) van der Waals (vdW) materials are burgeoning as one of the top candidates due to their various structural, miniaturized dimensionality, electronic engineering degrees of freedom, and they have highly tunable magnetic, electronic, and optical properties\cite{1,2,3,4,5,6}. The interlayer weak vdW interaction also unleashes the flexibility to affect physical properties by vdW stacking. Recently, sliding ferroelectricity originating from polar stacking of nonpolar monolayers transition metal dichalcogenides (TMDs)\cite{7,8,9}, and interlayer sliding in AlN and BN bilayers\cite{10,11}. Besides, stacking orders effectively regulate the magnetic ground state, resulting in the magnetic phase transition, such as the CrI$_3$\cite{12} and CrBr$_3$\cite{13} bilayers. More interestingly, the topological states can be tuned by stacking order in MnBr$_3$ bilayer\cite{14}. Through interlayer sliding of 2D materials, Ma $\emph{et al.}$ proved in theory that MnBi$_2$Te$_4$\cite{15}, Co$_2$CF$_2$\cite{16}, FeCl$_2$\cite{17}, etc., can realize layer-polarized anomalous Hall effect (LPAHE). Additionally, stacking order can effectively tune the intrinsic valley polarization of ferrovalley materials, such as the YI$_2$\cite{18} and VSiGeP$_4$\cite{19}. These findings suggest that the stacking order has a profound effect on determining the crystal symmetry, reallocating spin and charge between neighboring layers, and thus regulating the strength of ferroelectric, magnetic, topological, and valley polarization properties.

Multiferroics, materials exhibiting coupled two or more ferroic orders (i.e., ferromagnetic, ferroelectric, ferroelastic, ferrovalley), are especially interesting for the next generation electronic devices. At present, 2D multiferroic systems are often multiferroic heterostructures, such as Cr$_2$Ge$_2$Te$_6$/In$_2$Se$_3$\cite{20}, LaCl/In$_2$Se$_3$\cite{21}, and Cr$_2$COOH/Sc$_2$CO$_2$\cite{22}. However, the heterostructure system will increase the difficulty of fabrication device in an experiment, which is disadvantage to wider application. In contrast, if the 2D multiferroic can be realized by vdW stacking order in the same kind of material, it becomes more attractive. It will realize highly flexible and controllable coupling of the multiple ferroic orders. Therefore, the 2D multiferroic materials will further promote the application prospect in nanodevices.

In this work, we propose a new design of 2D multiferroics including magnetism, ferroelectric and ferrovalley. Here we focus on the coupling and phase transition of magnetic, ferroelectric, and ferrovalley by the effect of stacking orders in bilayer GdI$_2$. Firstly, we theoretically proposed that monolayer GdI$_2$ shows the spontaneously sizable valley polarization, which is manipulated by switching magnetization. Then, we found that bilayer GdI$_2$ endow the coexistence magnetism, ferroelectric and ferrovalley, supporting the designed target for manipulating magnetism and valley polarization via ferroelectric switching by interlayer sliding. Moreover, we reveal the microscopic mechanism of magnetic phase transition by spin Hamiltonian and electron hopping between layers. The highly flexible tunable multiferroic in bilayer GdI$_2$ offers a practical way for designing advanced valleytronic and spintronic devices on account of the couplings between multiferroic orders.

The 2D multiferroic systems with the coexistence of magnetism, ferroelectric, and ferrovalley are established. Practically, for a monolayer system with intrinsically spontaneous valley polarization, it can realize multiferroic by bilayer stacking and interlay sliding. As shown in Figure 1, it shows realizing the mechanism of 2D multiferroic with the coexistence of magnetism, ferroelectric, and ferrovalley. In AA stacking, the magnetic ground state is antiferromagnetic (AFM) coupling. Simultaneously, there are no spontaneous valley polarization and ferroelectric polarization in the system. When the AA stacking slide to AB or BA stacking, it will occur the magnetic phase transition from the AFM state to ferromagnetic (FM) state, and inducing the spontaneous valley polarization and ferroelectric polarization (signed by dark gray arrows). Moreover, the AB and BA stacking can be transformed into each other by interlayer sliding, leading to switch the ferroelectric polarization between +z and -z axis direction.

\begin{figure}[htb]
\begin{center}
\includegraphics[angle=0,width=1.0\linewidth]{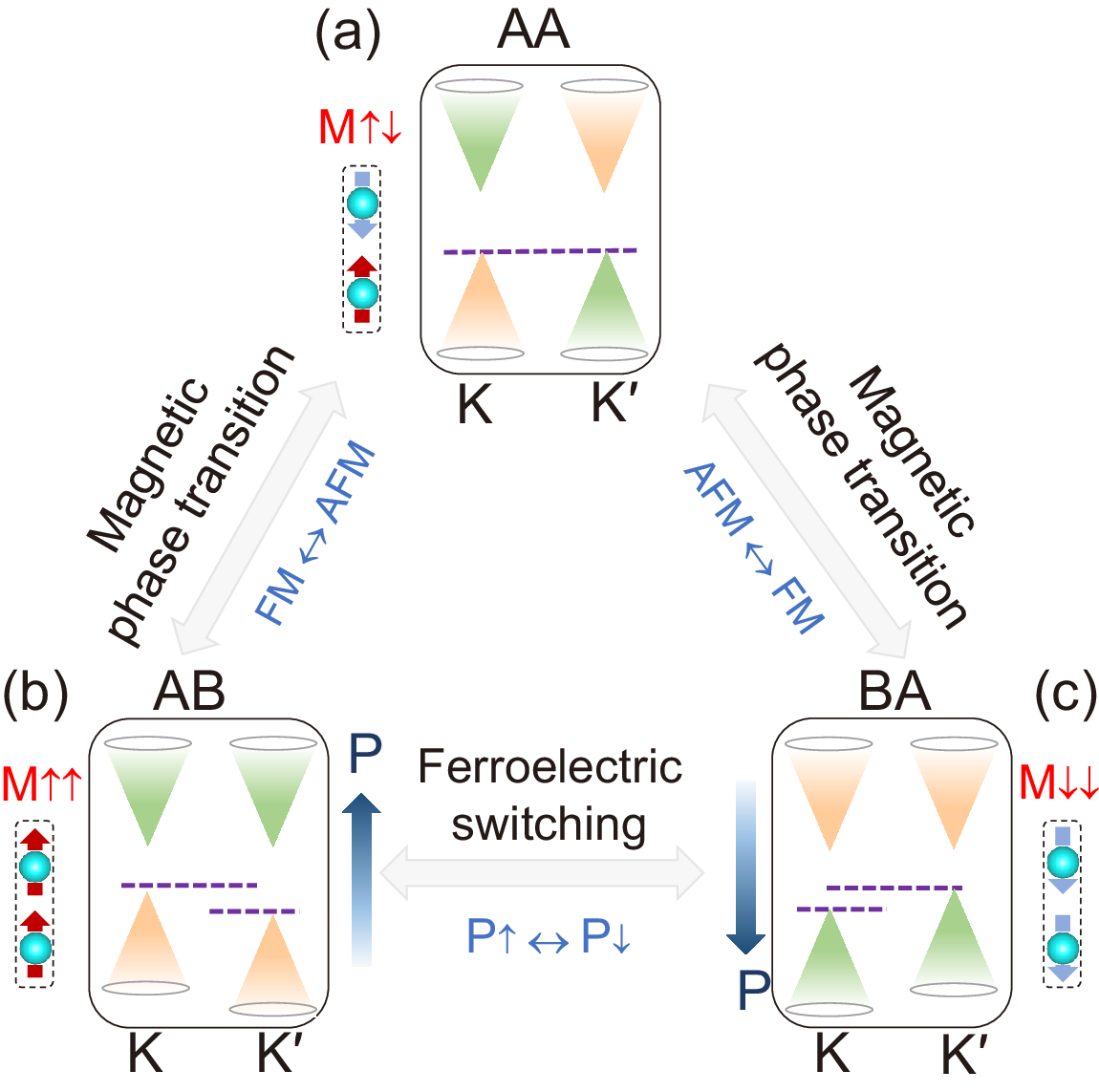}
\caption{Illustration of the mechanism of coexisting magnetism, ferroelectric, and ferrovalley multiferroic. (a) The magnetic ground state of AA stacking bilayer lattice is AFM state, and without spontaneous valley polarization. The AA stacking slide to AB stacking (b) or BA stacking (c), the magnetic ground state will switch to FM state. Simultaneously, spontaneous valley polarization will be realized. The valley polarization can also be manipulated via ferroelectric switching in AB stacking (b) and BA stacking (c). Orange and light green cones represent spin up and spin down bands, respectively. Dark gray arrows denote ferroelectric polarization $\emph{P}$.
}
\end{center}
\end{figure}

\begin{figure*}[htb]
\begin{center}
\includegraphics[angle=0,width=1.0\linewidth]{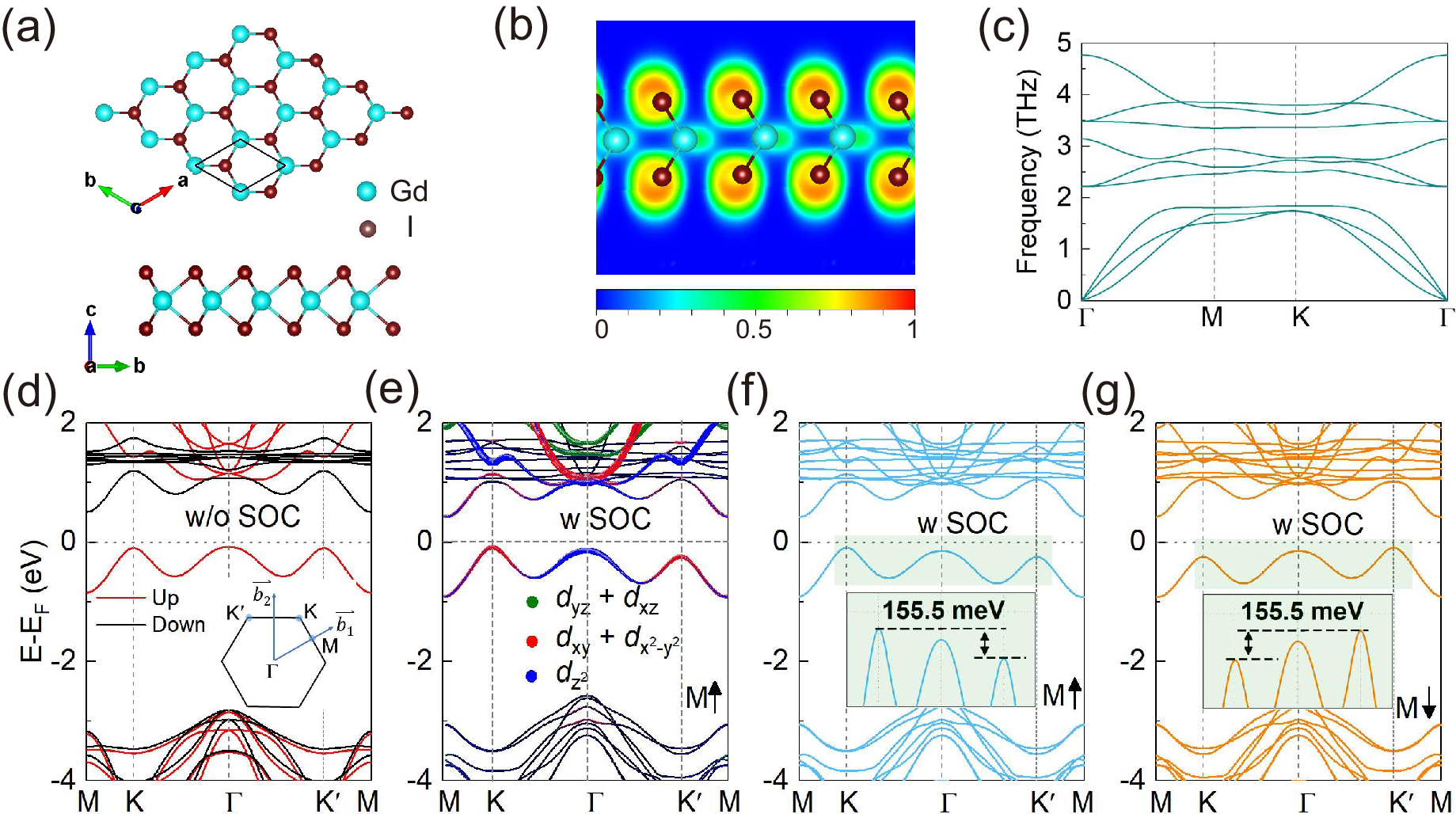}
\caption{(a) Crystal structures of monolayer GdI$_2$ from the top view and side view. (b) Electron localization function of monolayer GdI$_2$. (c) The calculated phonon spectrum along the high-symmetry. (d) Spin-polarized band structure of monolayer GdI$_2$. (e) The orbital-resolved band structure with considering SOC. (f, g) The band structures with considering SOC of magnetic moments along +z (f) and -z (g) axis directions, respectively.
}
\end{center}
\end{figure*}

As shown in Figure 2(a), it exhibits the crystal structure of monolayer GdI$_2$. The inversion symmetry is broken in monolayer GdI$_2$. The in-plane lattice constant for the fully optimized is 4.17 ${\rm \AA}$. Besides, in order to study the bonding characteristics of monolayer GdI$_2$, we calculate the electron localization function (ELF). As shown in Figure 2(b), the electrons are mainly localized around the Gd and I atoms, while those are negligible between the atoms, showing a typical ionic bonding for all the bonds. In addition, to evaluate the stability of monolayer GdI$_2$, we calculate the phonon dispersion spectrum. As shown in Figure 2(c), the absence of imaginary modes along the high-symmetry lines confirms the dynamical stability of monolayer GdI$_2$. Moreover, we calculate the formation energy that is -11.140 eV. It indicates that GdI$_2$ is easily prepared.

In order to make sure that the magnetic ground state of monolayer GdI$_2$, the rectangle supercell is constructed (see Figure S1). The energy difference $\Delta E$ = E$_{\rm FM}$ - E$_{\rm AFM}$ between FM and AFM state is found to be -266 meV, meaning that the magnetic ground state of monolayer GdI$_2$ is FM state. However, the successful preparation of 2D magnetic Cr$_2$Ge$_2$Te$_6$ \cite{3} and CrI$_3$ \cite{4} reveal that the magnetic anisotropy energy (MAE) plays an important role in the stability of magnetic order. Therefore, we investigate the MAE, it is defined as the energy difference MAE = E$_{001}$ - E$_{100}$ between the magnetic moment along [001] and [100]. The MAE is 0.777 meV per unit cell, standing for the magnetization along the x axis. For the monolayer GdI$_2$, the magnetic field (H = -$\frac{4}{3}\frac{K}{\mu_0M}$) of 3.58 T is needed to tune the direction of magnetization from the x to z axis.

Next, we concern on the band structures and associated valley properties of monolayer GdI$_2$. As shown in Figure 2(d), it is an indirect semiconductor in the absence of spin-orbit coupling (SOC), the valence band maximum (VBM) belong to the spin up band, located at the high symmetry K and K' points. While the conduction band minimum (CBM) is contributed by the spin down band, located at the high symmetry $\Gamma$ point. Note that the K and K' valleys of VBM are energetic degeneracy. When the SOC is switched on, as shown in Figure 2(e, f), the degeneracy between K and K' valleys are broken. The valley splitting is 155.5 meV in monolayer GdI$_2$, which is larger than VSe$_2$ (90 meV) \cite{36,37}, VSiXN$_4$ ($\thicksim$ 70 meV) \cite{38}, MoTe$_2$/EuO ($\thicksim$ 20 meV) \cite{39}, and other ferrovalley materials. More fascinatingly, when the magnetization direction is switched from the +z axis to -z axis, the valley polarization can be effectively tuned [see Figure 2(g)]. From the orbital-resolved band structure [see Figure 2(e) and Figure S2], the VBM bands are mainly dominated by Gd d$_{xy}$ and d$_{x^2-y^2}$ orbitals, while the CBM bands are contributed by Gd d$_{xy}$, d$_{x^2-y^2}$, and d$_{z^2}$ orbitals. Moreover, we calculate the valley polarization of U values from 2.5 eV to 5 eV. As shown in Figure S3, We can clearly observe that the valley polarization increases first and then decreases. However, the range of variation is very small only 0.5 meV.

\begin{figure*}[htb]
\begin{center}
\includegraphics[angle=0,width=1.0\linewidth]{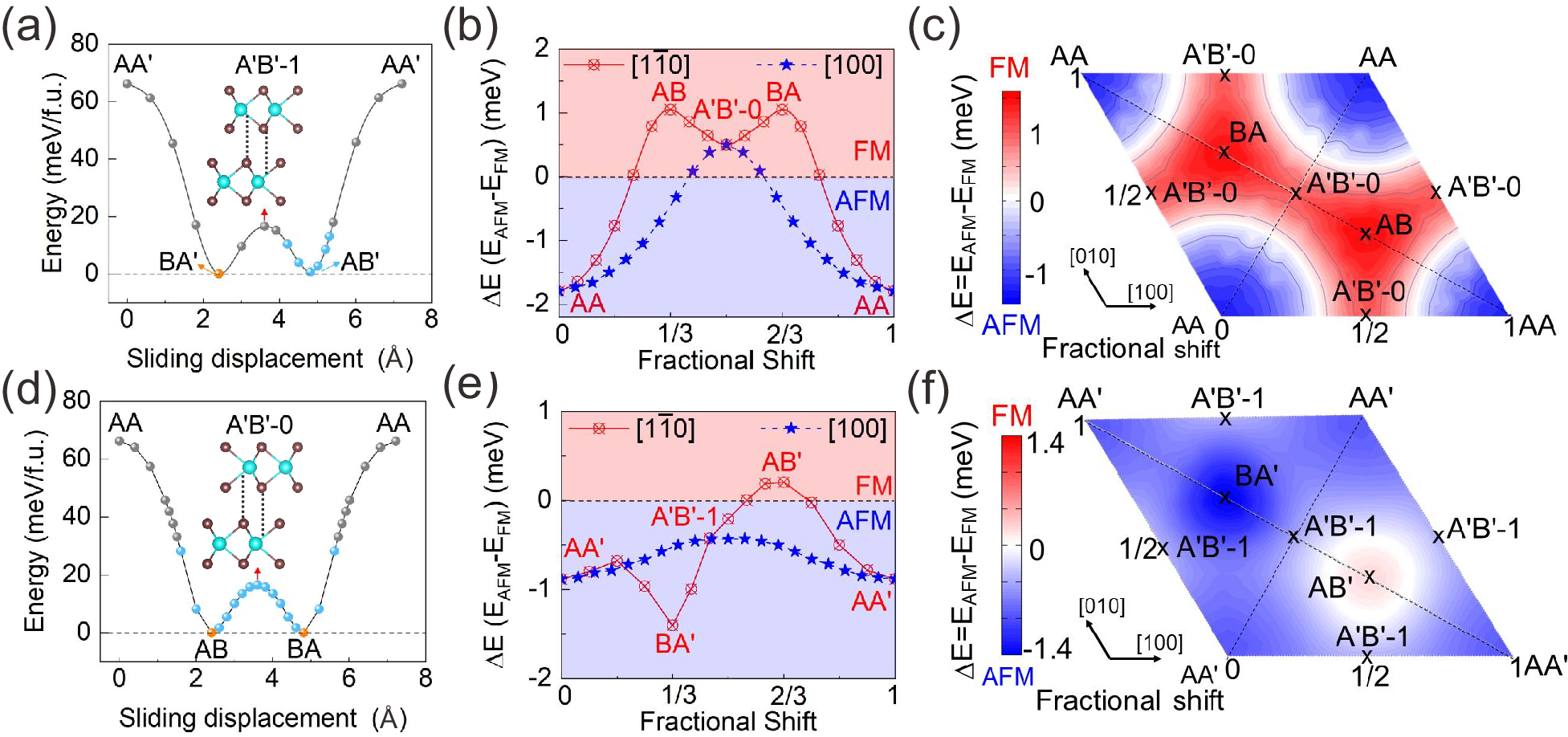}
\caption{ Sliding energy barrier between upper and lower layers of (a) AA stacking and (d) AA' stacking bilayer GdI$_2$. The gray and blue ball indicates that the magnetic ground state is AFM and FM state, respectively. The orange ball represents the energy minima. The energy difference between interlayer AFM and FM states as a function of interlayer translation along [1$\bar{1}$0] (solid red line) and [100] (dotted blue line) direction for AA stacking (b) and AA' stacking (e) bilayer GdI$_2$. The full space of lateral shifts for AA stacking (c) and AA' stacking (f) bilayer GdI$_2$. The positive or negative regions represent FM or AFM state, respectively.
}
\end{center}
\end{figure*}

To understand the microscopic mechanism in the valley splitting of valence bands, we use an effective Hamiltonian model to describe the physical nature of valley polarization induced by the SOC effect. Taking the SOC effect as the perturbation term,
\begin{equation}
\hat{H}_{SOC} = \lambda \hat{L} \cdot \hat{S} = \hat{H}_{SOC}^{0} + \hat{H}_{SOC}^{1},
\end{equation}
where $\hat{L}$ and $\hat{S}$ are orbital angular and spin angular operators, respectively. $\hat{H}_{SOC}^{0}$ denotes the interaction between the same spin states, while $\hat{H}_{SOC}^{1}$ represents the interaction between the opposite spin states. Since the VBM is contributed by spin up band, therefore, the term $\hat{H}_{SOC}^{1}$ can be neglected. For the $\hat{H}_{SOC}^{0}$, it can be written by the polar angles \cite{40}
\begin{equation}
\hat{H}_{SOC}^{0} = \lambda \hat{S}_{z'}(\hat{L}_zcos\theta + \frac{1}{2}\hat{L}_+e^{-i\phi}sin\theta + \frac{1}{2}\hat{L}_-e^{+i\phi}sin\theta),
\end{equation}
When the magnetization direction of monolayer GdI$_2$ along the out-of-plane, $\theta$ = $\phi$ = 0$^ \circ$, therefore, the $\hat{H}_{SOC}^{0}$ term can be simplified as
\begin{equation}
\hat{H}_{SOC}^{0} = \lambda \hat{S}_{z} \hat{L}_z,
\end{equation}

Considering  the C$_3$ symmetry and the contribution of orbital, we chose  $|$$\psi$$_v$$^{\tau}$$\rangle$=$\frac{1}{\sqrt{2}}$($|$d$_{xy}$$\rangle$+i$\tau$$|$d$_{x^2-y^2}$$\rangle$)$\otimes$$|$$\uparrow$$\rangle$, as the orbital basis functions for the VBM, where $\tau$ = $\pm$1 represent the valley index corresponding to $\rm K/\rm K'$. The energy of the K and K' valleys for the VBM can be written as E$_v$$^ \tau$ = $\langle$ $\psi$$_v$$^ \tau$ $|$ $\hat{H}$$_{SOC}^{0}$ $|$ $\psi$$_v$$^ \tau$ $\rangle$. Then, the valley splitting in the valence band can be written as
\begin{equation}
E_{v}^{K} - E_{v}^{K'} = i \langle d_{xy} | \hat{H}_{SOC}^{0} | d_{x^2-y^2} \rangle - i \langle d_{x^2-y^2} | \hat{H}_{SOC}^{0} | d_{xy} \rangle \approx 4\beta,
\end{equation}
where the $\hat{L}_z|d_{xy} \rangle$ = -2i$\hbar$$|d_{x^2-y^2} \rangle$, $\hat{L}_z|d_{x^2-y^2} \rangle$ = 2i$\hbar$$|d_{xy} \rangle$, and $\beta = \lambda \langle d_{x^2-y^2} |\hat{S}_{z'}| d_{x^2-y^2} \rangle$. The analytical result proves that the valley splitting of valence band is consistent with our DFT calculations ($E_{v}^{K}$ - $E_{v}^{K'}$ = 155.5 meV).

Layer stacking of 2D magnetic materials likely induces intriguing physical phenomena, which differ from the monolayer materials \cite{12,13,14,15,16,17,41}. Here, we investigate novel physical behaviors of bilayer GdI$_2$ by the stacking orders. Six typical stacking structures, AA, AB, BA, AA', AB', and BA', are shown in Figure S4. The AA stacking bilayer GdI$_2$ is constructed by primitively placing one layer on top of the other layer. Therefore, AA stacking layer GdI$_2$ has mirror symmetry between the two layers about the xy plane. AB and BA stacking are obtained by sliding t$_{1//}$[$\frac{2}{3}$, $\frac{1}{3}$, 0] and t$_{1//}$[$\frac{1}{3}$, $\frac{2}{3}$, 0] of AA stacking, respectively. However, the AA' stacking can be realized by rotating the upper layer 180$^{\circ}$ of AA stacking about the xy plane. AA' stacking could be transformed into the AB' and BA' stacking by by sliding t$_{1//}$[$\frac{2}{3}$, $\frac{1}{3}$, 0] and t$_{1//}$[$\frac{1}{3}$, $\frac{2}{3}$, 0], respectively.

\begin{table*}[htbp]
\caption{
The interlayer distance of nearest-neighboring and second-neighboring between Gd atoms, the number of magnetic exchange interactions per unit cell in square brackets, the calculated Heisenberg exchange parameters, and the energy difference between interlayer AFM and FM states for the different stacking configurations. }
\begin{tabular}{cccccccc}
	\hline
	Stacking   & Gd-Gd NN (${\rm \AA}$)   & Gd-Gd second-NN (${\rm \AA}$)    & J$_{1\bot}$ (meV)   & J$_{2\bot}$ (meV)   & J$_{\|}$ (meV)    & $\Delta E $ (meV)   \\
	\hline
	AA         & 8.139 [1]                & 9.145 [6]                        & 0.243               &  -0.003             &  -8.253           & -1.794               \\
	AB         & 7.888 [3]                & 8.922 [3]                        & -0.036              &  -0.007             &  -8.242           & 1.046              \\
	AB'        & 7.505 [1]                & 8.586 [6]                        &  0.002              &  -0.004             &  -8.235           & 0.206              \\
	BA'        & 7.915 [3]                & 8.946 [3]                        &  0.066              &  -0.008             &  -8.256           & -1.403               \\
	\hline
\end{tabular}
\end{table*}	

The sliding energy barriers of two kinds of stacking patterns are investigated to acquire the ground state stacking configurations. As shown in Figure 3(a), for the one kind stackings, AB and BA stackings exhibit the lowest energy, while the AA stacking has the highest energy. The transition AB stacking into BA stacking requires overcoming an energy barrier of 16.42 meV. For the other kinds of stacking [see Figure 3(d)], BA' stacking shows the lowest energy, while the energy of AB' stacking is 0.75 meV larger than that of BA' stacking. The above analysis shows that the AB, BA, and BA' are stable for bilayer GdI$_2$. To understand the effect of stacking on magnetism, we calculate the energy difference between the interlayer AFM and FM states, as shown in Figures 3(b) and 3(e) for the kinds of AA and AA' stackings, respectively. Seeing is believing, the magnitude of interlayer magnetic interaction for the kind of AA stacking is stronger than that of the kind of AA' stacking. Firstly, for the kind of AA stacking, we find that the AA stacking strongly prefers AFM, and the corresponding interlayer exchange energy is about -1.794 meV (see Figure S5). On the contrary, the interlayer exchange energy of AB and BA stackings is 1.046 meV (see Figure S5), it indicates that the magnetic phase transition occurs from AA stacking sliding to AB or BA stacking. Furthermore, for the other (the kind of AA' stacking), the vicinity of AB' stacking prefers to FM state, while the other regions favor AFM coupling. In a word, it indicates a strong coupling between the magnetism and stacking order.

Are there other intriguing stacking orders? In order to answer the question, we calculated the interlayer exchange energy of the kinds of AA and AA' stackings for the entire 2D space of lateral shifts. We chose a $6\times 6$ grid to employ the lateral shifts. Figure 3(c, f) shows the energy difference between interlayer AFM and FM states for the full 2D space of lateral shifts. One can see that the magnetic ground state switches between AFM and FM states as the interlayer stacking order changes. Moreover, the D$_{3h}$ symmetric AA stacking prefers the AFM state. However, the magnetic interlayer interaction become FM coupling for AB and BA stackings, when the D$_{3h}$ symmetric broken the xy mirror symmetry (C$_{3v}$). It is well known that the broken xy mirror symmetry for D$_{3h}$ point group induces the out-of-plane electric polarization $\emph{P}$ \cite{10}. As shown in Figure S6, the electrostatic potential difference with the positive (negative) discontinuity $\Delta \Phi$ = 22.86 (-22.86) meV is generated between the vacuum levels of upper and lower layers, which clearly proves that the spontaneous $\emph{P}$ produced along +z (-z) axis in the AB (BA) stacking bilayer. The electric polarization value of AB stacking bilayer GdI$_2$ is up to 3.68$\times$10$^{-12}$ C/m. The AB and BA stacking bilayers can realize reciprocal transformation by interlayer sliding, therefore, we came to the conclusion that the sliding ferroelectric exists in bilayer GdI$_2$, which can also be confirmed from the general theory of bilayer sliding ferroelectricity \cite{42}. It indicates that the magnetic ground state can effectively be tuned by the sliding ferroelectric. Note that the kind of AA' stacking inexist spontaneous polarization due to the 180$^\circ$ rotation between two layers induced the out-of-plane polarization to cancel (see Figure S7).

To understand the microscopic mechanism of magnetic coupling, we consider the simple spin Hamiltonian
\begin{equation}
H=E_0+\sum_{i,j}J_{\|}S_i\cdot S_j+\sum_{i,k}J_{1\bot}S_i\cdot S_k+\sum_{i,l}J_{2\bot}S_i\cdot S_l,
\end{equation}
where $E_0$ is the ground state energy independent of the spin configurations. \emph{S}$_i$, \emph{S}$_j$, \emph{S}$_k$, and \emph{S}$_l$ denote the magnetic moments at sites $\emph{i}$, $\emph{j}$, $\emph{k}$, and $\emph{l}$, respectively. \emph{J}$_{\|}$, \emph{J}$_{1\bot}$, and \emph{J}$_{2\bot}$ represent the intralayer, nearest-neighbor (NN) interlayer, and second-neighbor (second-NN) interlayer Gd-Gd exchange interactions, respectively (see Figure 4). The details for the calculation of Heisenberg exchange parameters based on the above spin Hamiltonian and total energy calculations employing DFT are shown in the Supporting Information and Table I. The intralayer exchange is strongly FM coupling ($\thicksim$ -8.25 meV), while the interlayer exchange interaction depends on the stacking order.

\begin{figure*}[htb]
\begin{center}
\includegraphics[angle=0,width=1.0\linewidth]{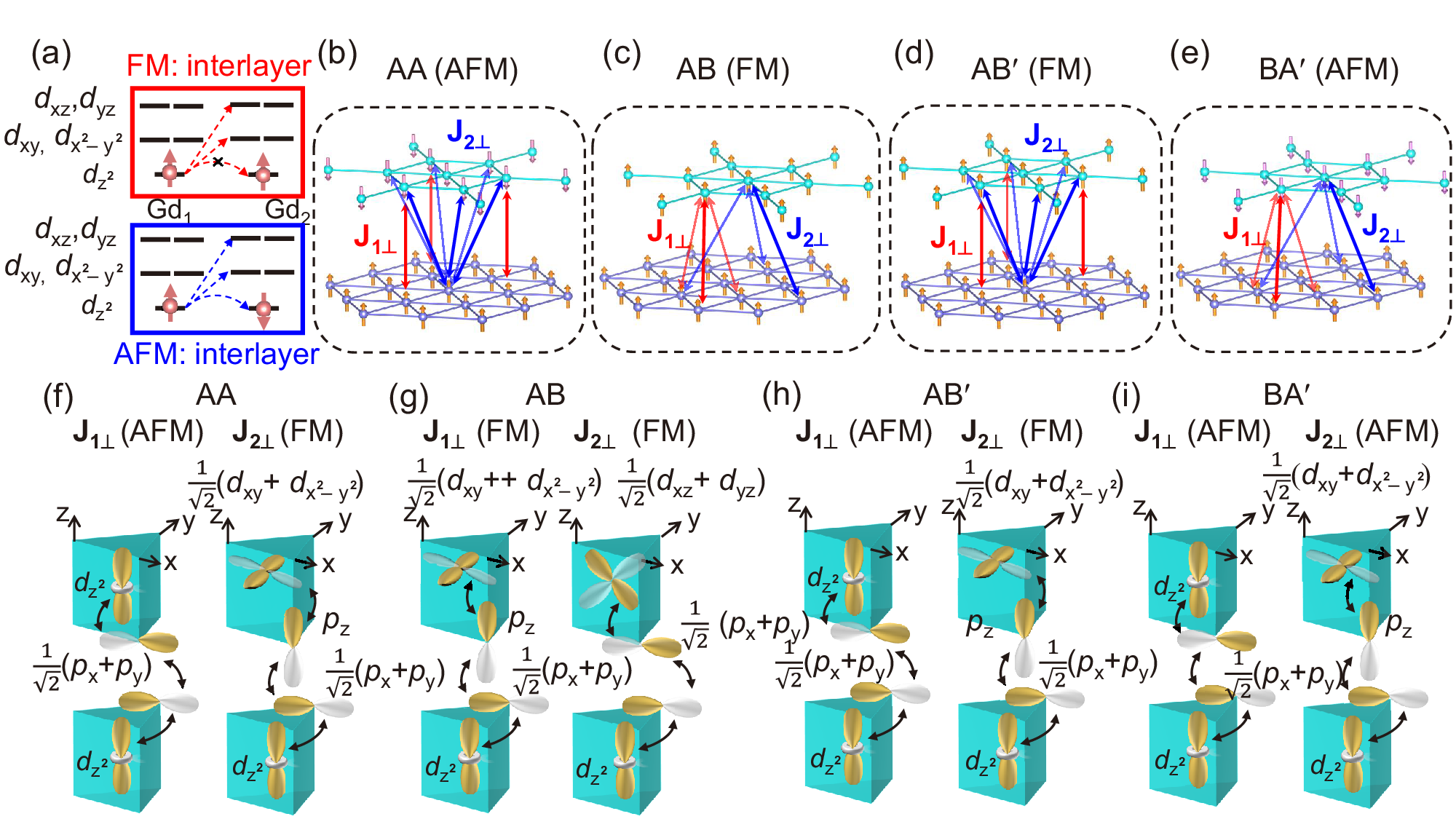}
\caption{(a) Schematic of the orbital dependent interlayer exchange interactions. The hopping from d$_{z^2}$-d$_{z^2}$ is allowed in the AFM exchange, whereas it is prohibited in the FM exchange. (b-e) The interlayer Gd nearest neighbor and second neighbor corresponds to J$_{1\bot}$ (red) and J$_{2\bot}$ (blue) for AA (b), AB (c), AB' (d), and BA' (e) stacking, respectively. (f-i) The schematics of super-superexchange for AA (f), AB (g), AB' (h), and BA' (i) stacking.
}
\end{center}
\end{figure*}

Since the Gd atoms in GdI$_2$ are in a 4f$^7$5d$^1$ electronic configuration (see Figure S2), according to the Pauli exclusion principle and Hund's rule, the electron configuration will half fill the f and d$_{z^2}$. In the trigonal prismatic crystal field, the schematic illustrations of Gd-5d and Gd-4f orbital splitting is shown in Figure S2(b). According to the energy minimizing principle of hopping, the magnetic exchange between the two layers is determined by the d orbital hopping. As shown in Figure 4(a), it exhibits the schematic of the FM and AFM exchange interactions between Gd atoms in the interlayer. The hopping between d$_{z^2}$ and d$_{z^2}$ is prohibited for FM spin configuration (red), while it is allowed for AFM spin configuration (blue). Hence, d$_{z^2}$-d$_{z^2}$ hybridization results in AFM. On the other hand, hopping of the from d$_{z^2}$-d$_{xy}$/d$_{x^2-y^2}$ and d$_{z^2}$-d$_{xz}$/d$_{yz}$ gives rise to an exchange coupling that is overwhelmingly FM from the viewpoint of the local Hund coupling \cite{43,44}. The interlayer Gd-Gd exchange interactions of all stacking are mediated via the hybridization between the I-p orbitals. Therefore, the nature of interlayer exchange interaction is super-superexchange. The magnetism of stacking order dependence derives from the competition between the orbital hybridizations of different interlayer.

In the following, we take four typical stacking as examples for detailed analysis. For the AA stacking, Figure 4(b) shows the interlayer NN exchange interaction of Gd-Gd J$_{1\bot}$ (red) and second-NN exchange interaction of Gd-Gd J$_{2\bot}$ (blue) for AA stacking. As shown in Figure 4(f), J$_{1\bot}$ is predominated via virtual excitations between Gd half-filled d$_{z2}$ orbitals, causing an AFM coupling \cite{45}. By contrast, J$_{2\bot}$ is dominated by virtual excitations between Gd half-filled d$_{z^2}$ and the empty d$_{xy}$/d$_{x^2-y^2}$ orbitals, leading to the FM coupling. As listed in Table I, although AA stacking has only one J$_{1\bot}$ bond and six J$_{2\bot}$ bonds per unit cell, J$_{1\bot}$ is much larger than J$_{2\bot}$. Therefore, the NN interlayer AFM coupling dominates the second-NN interlayer FM coupling, resulting in the AFM state for AA stacking GdI$_2$. For the AB stacking, the sliding one layer t$_{1//}$[$\frac{2}{3}$, $\frac{1}{3}$, 0] with relative to the other breaks the interlayer hybridization between I-p electrons and produces new ones. As shown in Figure 4(c), the NN ligancy number is increased compared with the AA stacking the second-NN ligancy number is decreased (see Table I). Simultaneously, the virtual excitations of J$_{1\bot}$ become between Gd half-filled d$_{z^2}$ and the empty d$_{xy}$/d$_{x^2-y^2}$ orbitals, indicating the FM coupling. While J$_{2\bot}$ is dominated by virtual excitations between Gd half-filled d$_{z^2}$ and the empty d$_{xz}$/d$_{yz}$ orbitals, contributing to also the FM coupling. As a result, the magnetic ground state becomes FM state. Although the coordination number of AB' (BA') stacking is consistent with AA (AB) stacking, they exhibit opposite magnetic ground states. It originates from the symmetry difference [see Figure S4, Figure 4(b-i), and Table I].

Stacking orders not only induces the changes of magnetic ground state and ferroelectric polarization, the valley polarization is also transformed due to the different symmetries. As shown in Figure 5(a) and Figure S8(a-d), the band structures of AA stacking bilayer GdI$_2$ are investigated based on the AFM (M$\uparrow \downarrow$) magnetic ground state. In the absence of SOC, spin up and spin down bands are degenerate [see Figure S8(a)]. When the SOC is switched on, as shown in Figure 5(a), the degeneracy of spin up and spin down bands disappear. In detail, we focus on the valence band. The spin up band shifts above the spin down band in K valley, while the spin up band shifts below the spin down band in K' valley. More interestingly, the spin up and spin down bands of K and K' valleys are degenerate in energy. From the orbital-resolved band structure [see Figure S8(b)], the VBM bands are also dominated by Gd d$_{xy}$ and d$_{x^2-y^2}$ orbitals, it is consistent with monolayer GdI$_2$. Moreover, as shown in Figure 5(b), the Berry curvatures of K and K' valleys have equal magnitudes and opposite signs (26.2 ${\rm \AA}^2$ for K point and -26.2 ${\rm \AA}^2$ for K' point).

\begin{figure}[htb]
\begin{center}
\includegraphics[angle=0,width=1.0\linewidth]{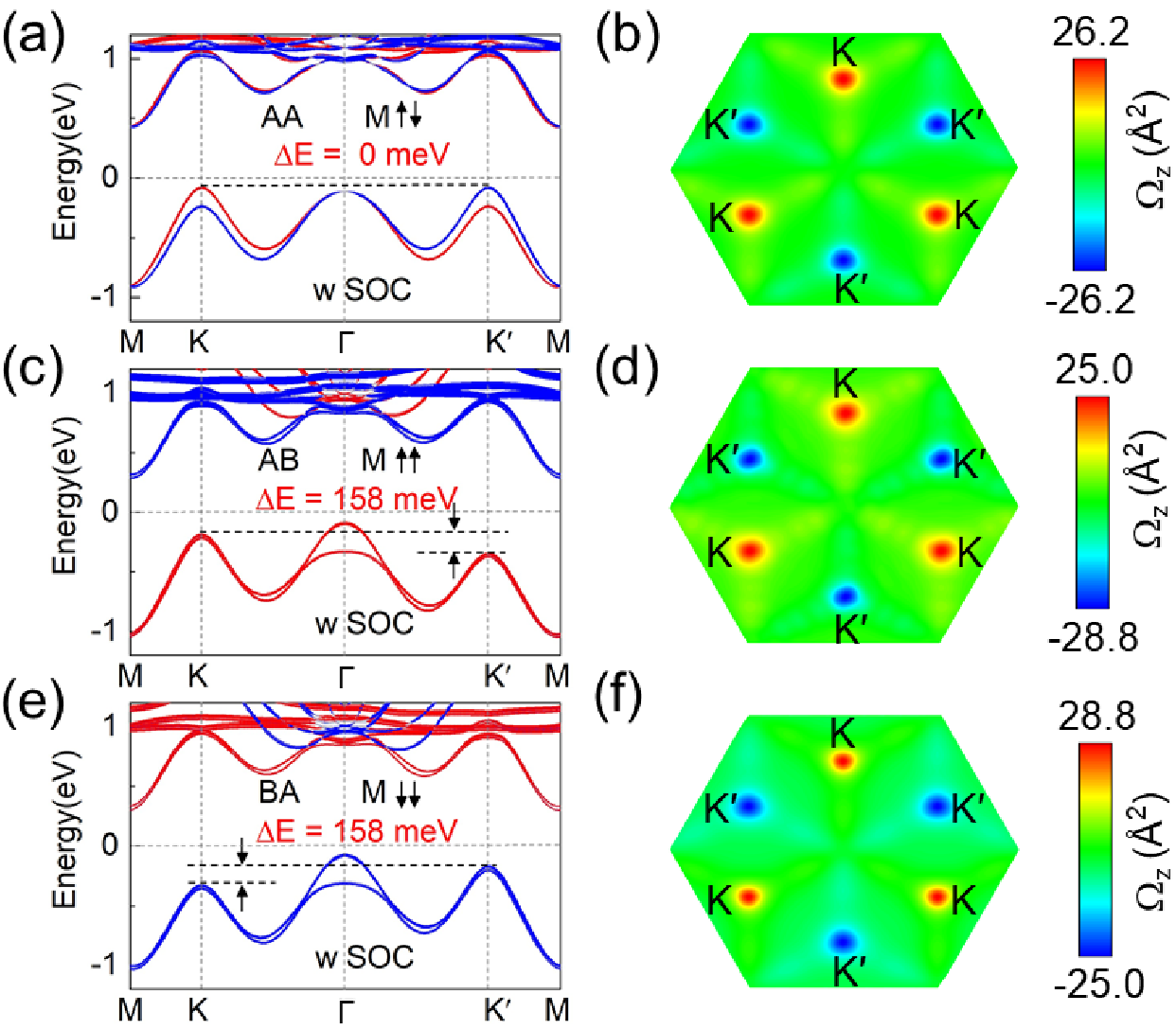}
\caption{ Spin-resolved band structures and Berry curvatures of AA (a, b), AB (c, d), and BA (e, f) stacking bilayer GdI$_2$ with considering SOC effect.
}
\end{center}
\end{figure}

When the AA stacking slides to AB stacking, the degeneracy of spin up and spin down bands disappear in the absence of SOC [see Figure S8(e)]. It originates from the magnetic ground state from AFM state transition to FM state. When the SOC is included, as shown in Figure 5(c), it produces a valley polarization of 158 meV. It indicates that the interlayer sliding induces the out-of-plane ferroelectric polarization, simultaneously, leading to the magnetic phase transition and valley polarization. Due to the valley polarization generating, as shown in Figure 5(d), the Berry curvature magnitudes of K and K' valleys are no longer equal, and the signs are still opposite (25.0 ${\rm \AA}^2$ for K point and -28.8 ${\rm \AA}^2$ for K' point). When further slide to BA stacking, the valley polarization is reversed by the ferroelectric switching, as shown in Figure 5(e). Meanwhile, the Berry curvatures of K and K' valleys change to 28.8 ${\rm \AA}^2$ and -25.0 ${\rm \AA}^2$ [see Figure 5(f)]. In addition, the other stacking also produces intriguing coupling of magnetic, ferroelectric, and ferrovalley orders (see Figure S9).

In conclusion, we propose a mechanism to realize coexisting magnetism, ferroelectric, and ferrovalley multiferroic in 2D materials. The mechanism is proved in bilayer GdI$_2$. By transforming the interlayer stacking order, one can regulate the magnetic ground state, ferroelectric polarization, and valley polarization for bilayer GdI$_2$. The magnetic ground state of AA stacking is AFM state with the degeneracy at the VBM of K and K' valleys. As the AA stacking slides to AB (BA) stacking, its spontaneous valley polarization occurs due to the magnetic ground state transforming from AFM to FM state. The AB stacking changes to the BA stacking state by interlayer sliding, and the ferroelectric polarization and valley polarization are switched. Moreover, we reveal the microscopic mechanism of magnetic phase transition by spin Hamiltonian and electron hopping between layers. Our work offers not only a novel 2D multiferroic material but also an efficient means to tune magnetic, ferroelectric, and ferrovalley properties.

\subsection{AUTHOR INFORMATION}
\textbf{Corresponding Authors}\\
pli@xjtu.edu.cn
\\
\textbf{Notes}\\
The authors declare no competing financial interest.
\\

\begin{acknowledgement}
This work is supported by the National Natural Science Foundation of China (Grant No. 12004295). P. Li thanks China's Postdoctoral Science Foundation funded project (Grant No. 2022M722547), the Fundamental Research Funds for the Central Universities (xxj03202205), and the Open Project of State Key Laboratory of Surface Physics (No. KF2022$\_$09).
\end{acknowledgement}

\begin{suppinfo}
Details of the calculation methods, spin Hamiltonian model, spin charge density, orbital-resolved band structures, the effect of U value on valley polarization, stacking structures of bilayer GdI$_2$, the energy difference between interlayer AFM and FM for bilayer, the plane averaged electrostatic potential, and band structures for bilayer GdI$_2$.
\end{suppinfo}

\newpage
\section*{Graphical TOC Entry}
\begin{center}
\includegraphics[angle=0,width=1.0\textwidth]{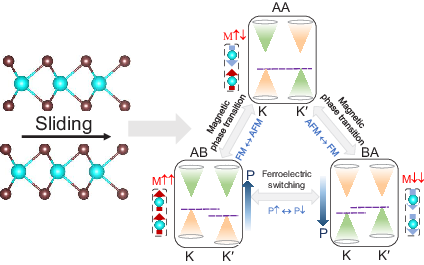}
\end{center}

\end{document}